\documentclass[english,prb,preprint,superscriptaddress,showpacs,preprintnumbers,amsmath,amssymb]{revtex4}
\usepackage[T1]{fontenc}
\usepackage[latin1]{inputenc}
\usepackage{graphicx}

\makeatletter

\makeatletter \tolerance = 10000 \tolerance = 10000

\usepackage{color}

\makeatother

\usepackage{babel}
\makeatother
\begin{document}

\title{Switching mechanism of photochromic diarylethene derivatives molecular junctions}

\author{Jing Huang}
\affiliation{Hefei National Laboratory for Physical Sciences at
Microscale, University of Science and Technology of China, Hefei,
Anhui 230026, People's Republic of China}

\author{Qunxiang Li}\thanks{Corresponding author. E-mail: liqun@ustc.edu.cn}
\affiliation{Hefei National Laboratory for Physical Sciences at
Microscale, University of Science and Technology of China, Hefei,
Anhui 230026, People's Republic of China}

\author{Hao Ren}
\affiliation{Hefei National Laboratory for Physical Sciences at
Microscale, University of Science and Technology of China, Hefei,
Anhui 230026, People's Republic of China}

\author{Haibin Su}
\affiliation{Division of Materials Science, Nanyang
Technological University, 50 Nanyang Avenue, 639798, Singapore}

\author{Q.W.Shi}
\affiliation{Hefei National Laboratory for Physical Sciences at
Microscale, University of Science and Technology of China, Hefei,
Anhui 230026, People's Republic of China}

\author{Jinlong Yang}\thanks{Corresponding author. E-mail: jlyang@ustc.edu.cn}
\affiliation{Hefei National Laboratory for Physical Sciences at
Microscale, University of Science and Technology of China, Hefei,
Anhui 230026, People's Republic of China}

\date{\today}

\begin{abstract}
The electronic transport properties and switching mechanism of
single photochromic diarylethene derivatives sandwiched between two
gold surfaces with closed and open configurations are investigated
by a fully self-consistent nonequilibrium Green's function method
combined with density functional theory. The calculated transmission
spectra of two configurations are strikingly distinctive. The open
form lacks any significant transmission peak within a wide energy
window, while the closed structure has two significant transmission
peaks on the both sides of the Fermi level. The electronic transport
properties of the molecular junction with closed structure under a
small bias voltage are mainly determined by the tail of the
transmission peak contributed unusually by the perturbed lowest
perturbed unoccupied molecular orbital. The calculated on-off ratio
of currents between the closed and open configurations is about two
orders of magnitude, which reproduces the essential features of the
experimental measured results. Moreover, we find that the switching
behavior within a wide bias voltage window is extremely robust to
both substituting F or S for H or O and varying end anchoring atoms
from S to Se and Te.
\end{abstract}


\maketitle

\section{INTRODUCTION}

A critical mission of the molecular electronics is to develop
innovative devices at single molecular scale. The representative
molecular wires, rectifiers, switches, and transistors have been
intensively studied in the past years.\cite{BookRev,Nature00aviram}
Obviously, a single molecular switch holds great promise since the
switch is a crucial element of any modern design of memory and logic
applications. Now various schemes have been proposed to realize
molecular switching process including relative motion of
molecule internal structure,\cite{PRL06young,AC06-revers,Science99-JChen,APL02,NM05amy}
change of molecule charge states,\cite{JACS00,JACS02} and bond
fluctuation between the molecule and their electrical
contacts.\cite{Science03-bond} Recently, an alternative routine has
been suggested to design switches based on single stably existing
molecule which can reversibly transform between two conductive
states in response to external
triggers.\cite{CCRev04Tian,JPPC04Kenji,PRL04zhang,JCP06jing} Among
various triggers, light is a very attractive external stimulus
because of the ease of addressability, fast response times, and
compatibility with a wide range of condensed phases.

The switching properties through the so-called photochromic
molecules have been carried out by several experimental and
theoretical
groups.\cite{EJIC00,PRL03diana,Nano05jin,Adv06nathalie,Chem06tibor,PRL04jun,PRB05min,CPL05masakazu}
In particular, the dithienylcyclopentene (DTC) derivatives (as the
central switching unit) hold great promise as artificial
photoelectronic switching molecules because of their reversible
photo-induced transformations that modulate electrical
conductivity and their exceptional thermal stability and fatigue
resistance.\cite{EJIC00,PRL03diana,Nano05jin,Adv06nathalie,Chem06tibor}
Using the mechanically controllable break-junction technique,
Duli$\acute{c}$ \emph{et al.}\cite{PRL03diana} designed molecular
switches based on DTC molecules
(1,2-bis{[}5$^{\prime}$-(5$^{\prime\prime}$-acetylsulfanylthien-2$^{\prime\prime}$-
yl)-2$^{\prime\prime}$-methylthien-3$^{\prime}$-yl] cyclopentene)
with two thiophene linkers, however, which operates only one-way,
i.e. from conducting to the insulating state under visible light
with $\lambda$=546 nm with resistance change at 2-3 orders of
magnitude. Interestingly, He \emph{et al.}\cite{Nano05jin} found
that the transition can be optical two-way for DTC molecules where
H atoms in cyclopentene are substituted by six F atoms
(fluorined-DTC). The single-molecule resistance in the open form
is about 130 times larger than that of in the closed structure
measured by using scanning tunneling microscopy with a gold tip.
In a parallel study, Katsonis \emph{et al.}\cite{Adv06nathalie}
used aromatic (meta-phenyl) linkers and observed that the
light-controlled switching of single DTC molecules connected to
gold nanoparticles was reversible. Very recently, to improve the
poor stability of such kind of conjugated molecules with thiols on
both ends, Taniguchi \emph{et al.}\cite{JACS06masateru} developed
an interconnect method in solution for diarylethene photochromic
molecular switches that can ameliorate electrode-molecule binding,
molecular orientation, and device functions. In their experiment,
one light-controlled switching molecule consists of a central
fluorined-DTC molecule, diaryls on two sides and two thiol groups
at both ends. The corresponding closed and open forms are shown
Figure 1(a). Consequently, the current through the molecular
junction with closed structure is about 20 times larger than that
of the open form measured by STM.\cite{JACS06masateru}

To gain better understanding of these experimental observations,
several theoretical work have been carried
out.\cite{PRL04jun,PRB05min,CPL05masakazu} Li \emph{et
al.}\cite{PRL04jun} performed quantum molecular dynamics and
density functional theory (DFT) calculations on the electronic
structures and transport properties through several photochromic
molecules with several different spacers sandwiched between gold
contacts. They chose dithienylethene (DTE) derivatives to model
the experimental measured molecules and predicted an about 30
times conduction enhancement when converting the open form into a
closed one by optical technique. Subsequently, two research groups
independently investigated the switching properties of DTC
molecular junctions, and found that the transmission peak
originates from the highest occupied molecular orbital (HOMO) of
the closed form lying near the electrode Fermi
level.\cite{PRB05min,CPL05masakazu} In their calculations, the
electrodes are simulated with cluster models and the effects on
the transport properties coming from six H hydrogen atoms
substituted by F atom are not considered. Till now, to our best
knowledge, there is no theoretical study about the switching
mechanism through exactly the same measured molecules
(fluorined-DTC with two diaryls on two sides, named diarylethene
in Taniguchi \emph{et al.}'s experiment). In this paper, we employ
the non-equilibrium green's function technique (NEGF) combined
with DFT method to address electronic transport and switching
behavior of diarylethene based molecular junction. Moreover, we
examine the robustness of this type of switching device against
various chemical substitution (where six F atoms in the peripheral
of cyclopentene and S atoms in thienyl are substituted by H and O
atoms, respectively) and alternations of anchoring atoms.

\section{COMPUTATIONAL MODEL AND METHOD}

The computational model system is schematically illustrated in Fig. 1(b). The
molecules with open and closed configurations are sandwiched between
two gold electrodes through S-Au bonds. The Au (111) surface is
represented by a (4$\times$4) cell with periodic boundary
conditions. Since the hollow site configuration is 
energetically preferable by 0.2 and 0.6 eV than the bridge and atop
sites, respectively,\cite{adsorbed} the diarylethene molecule connects to sulfur
atoms which are located at hollow sites of two Au (111) surfaces.
Both electrodes are repeated by three layers (A, B, and C). The
whole system is arranged as (BCA)-(BC-molecule-CBA)-(CBA), which can
be divided into three regions including the left lead (BCA), the
scattering region, and the right lead (CBA). The scattering regions
include a diarylethene molecule, two surface layers of the left
(BC), and three surface layers of the right lead (CBA), where all
the screening effects are included into the contact region, within
which the charge-density matrix is solved self-consistently with the
NEGF method.

The electronic transport properties are studied by the NEGF combined
with DFT calculations, which are implemented in ATK
package.\cite{Atk} This methodology has been adopted to explain
various experimental results
successfully.\cite{USTCworks,PRL02Nielsen} In our calculations,
Ceperley-Alder local-density approximation is
used.\cite{PRL80-CALDA} Core electrons are modeled with
Troullier-Matrins nonlocal pseudopotential, and valence electrons
are expanded in a SIESTA localized basis set.\cite{Siesta,Supply} A
energy cutoff of 150 Ry for the grid integration is set to present
the accurate charge density. The optimized electrode-electrode
distance is 39.5 {\AA} for the closed configuration which is 0.7
{\AA} longer than that of the open one. All atomic positions are
relaxed and the corresponding gold-sulfur distance is 2.5 {\AA},
which is close to the typical theoretical values.\cite{CMS03stokbro}
In addition, we find that the geometric changes of two diarylethene
molecule sandwiched between two Au(111) surfaces are negligible
comparing to the corresponding free molecules.

\section{RESULTS AND DISCUSSION}

\subsection{Electronic structures of free diarylethene molecules with closed
and open structures}

Atomic positions of two free diarylethene molecules with closed
and open structures are optimized by Gaussion03 package at general
gradient approximation level.\cite{G03} In the ground electronic
states, both optimized configurations are featured by out-of-plane
distortions. The central dihedral angle is 60 degrees for the open
form, while only about 8 degrees for the closed one. This
distortion leads the distance between carbon and carbon bond (the
bond can be broken by photon) close to 4.0 {\AA} in the open case
compared to 1.5 {\AA} for the closed configuration. These
important geometric parameters are consistent with the previous
DFT predictions for bisbenzothienylethene
molecules.\cite{JPCA06clark} Experimental studies have
demonstrated that the molecule can transform reversibly between
the closed and open forms by shining ultraviolet and visible
lights, respectively.\cite{JACS06masateru}

Drawing from the chemical intuition, one would expect that the
electronic structures have distinctive characteristics due to the
significant geometric difference between closed and open structures.
For example, it is clear that both single and double bonds appearing
in the central switching unit get almost swapped within the closed
and open configurations as shown in Fig. 1(a). The number of double
bonds is 9 in the open form in contrast to 8 in the closed one.
Thus, the energy of HOMO in the open form is expected to be lower
than that in the closed one.\cite{PRB05min} In deed, the energies of
the HOMO and lowest unoccupied molecular orbital (LUMO) of the
closed form are -4.6 and -3.3 eV, respectively, whereas the HOMO and
LUMO energies of the open one are -4.9 and -2.7 eV. The frontier
orbital localizes primarily on each conjugated unit of the molecule
or on the central switching unit for the diarylethene molecule with
open configuration. The molecule in the closed form belongs to a
conjugated system, whose HOMO and LOMO orbitals are essentially
delocalizated $\pi$ orbitals extending over the entire molecule.
More interesting, when six F atoms in the peripheral of cyclopentene
are substituted by H atoms, we find that the HOMO and LUMO energies
of this modified molecule shift dramatically to -4.2 and -2.5 eV for
the closed form, and to -4.7 and -1.6 eV for the open one
respectively. These remarkable differences of the geometries and
electronic structures are expected to affect significantly transport
properties.

\subsection{Transport properties of diarylethene molecular junctions}

The currents through the molecular junction with closed and open
configurations in the bias voltage range {[}-1.0, 1.0V] are
calculated by the Landauer-B\"{u}tiker formalism.\cite{Butt} It
should be pointed out that at each bias voltage, the current is
determined self-consistently under the nonequilibrium condition. The
calculated I-V curves are presented in Figure 2. The triangles
linking with black solid lines are for the diarylethene molecular
junction, while the circles linking with short red dotted lines
stand for the junction where six F atoms in the peripheral of
cyclopentene are substituted by H atoms. The filled (empty) symbols
correspond to the closed (open) structures. Our calculations capture
the key features of the experimental results.\cite{JACS06masateru}
The current through the closed form is remarkably higher than that
of the open one. When the diarylethene molecule in the junction
changes from a closed configuration to the open one, the molecular
wire is predicted to switch from the \emph{on} (low resistance)
state to the \emph{off} (high resistance) state. The current
enhancement is quantified by the on-off ratio of current defined as
$R(V)=I_{closed}(V)/I_{open}(V)$. For example, the current of the
closed form at 1.0 V is about 4.5 $\mu$A, which is about 500 times
larger than that of the open case. Such a large on-off ratio in this
given range of bias voltage can be readily measured and is desirable
for the real application. Note that the predicted on-off ratio at
1.0 V is larger by about one order of magnitude compared to
experiment.\cite{JACS06masateru} We think one possible reason for
this discrepancy is the limitation of the computational method. It
is well known that the calculated value of the current through
molecular junction using NEGF combined with DFT is larger about 1-2
orders of magnitude than that of these experimental measured
result.\cite{CMS03stokbro,USTCworks} Other two possible reasons are
environment effect and geometry difference. Firstly, solvent effect
is not considered in presented calculations. Secondly, in our
computational model, diarylethene molecules are directly bound to
gold electrodes through Au-S bonds in vacuum. In the experimental
setup, the central switching molecules bind to the long orientation
control molecules (polyrotaxane), which connect to the interface
control molecules (4-iodobenzenethiol) anchored with gold
nanoelectrodes in solution (the distance between two electrodes is
about 30 nm).\cite{JACS06masateru} Note that the slight geometric
distortion due to the molecule-electrode interaction can result in a
slight asymmetry in the calculated I-V curves at small bias voltage
range as shown in the inset below right of Fig. 2 in small scale for
clarity.

To understand the dramatic difference in conductivities of the
closed and open configurations, we compute the energy dependence of
total zero-bias voltage transmission spectra shown in Figure 3,
where the Fermi level (E$_{F}$) is set to be zero for clarity.
Generally speaking, the conductance of the molecular junction is
determined by the number of the eigenchannel, the properties of the
perturbed frontier orbitals of the molecule due to the presence of
the gold electrodes and the alignment of the metal Fermi level
within the perturbed HOMO-LUMO gap.\cite{Atk} Applying an effective
scheme named molecular projected self-consistent Hamiltonian (MPSH)
method,\cite{Atk} the orbital energies and eigenstates (referred as
perturbed MOs) of the MPSH are obtained and plotted in Fig. 3. The
energy positions of these perturbed MOs relative to the E$_{F}$ are
denoted in Figs. 3(a) and 3(b) with red short vertical lines, which
match nicely with the transmission peaks. The spatial distributions
of the perturbed-HOMOs and -LUMOs are presented in Fig. 3 locating
on the right and left sides of the E$_{F}$, respectively. Both
calculated conductances are very small at zero bias. It is
4.2$\times$10$^{-2}$ G$_{0}$ (G$_{0}$=2e$^{2}$/h) for the closed
configuration at the E$_{F}$, and only 5.4$\times$10$^{-5}$ G$_{0}$
for the open one which is about 800 times smaller than the former
one. The diarylethene molecule with a closed structure has two broad
and strong transmission peaks locating at -0.8 and 0.5 eV,
respectively. For the open form, note that the lack of any
significant peaks in between -1.5 and 1.7 eV clearly elucidates its
lower conductivity.

More importantly, the transmission spectra display extraordinarily
discrepant characteristics. It is clear that for the diarylethene
molecular junction with closed structure, the significant
transmission peaks locating below and above the E$_{F}$ (about -0.8
and 0.5 eV) are mainly contributed by the perturbed-HOMO and -LUMO,
respectively. Notably, the perturbed HOMOs and LUMOs of the closed
configuration in Fig. 3(a) are delocalized $\pi$-conjugated
orbitals, which provide good channels for electron tunneling through
the molecular junction and lead to two significant transmission
peaks. Very interestingly, the transport properties are predominated
by the tail of the perturbed LUMO contributed transmission peak at
small bias voltage (for example, less than 1.0 V), since the
transmission coming from the perturbed LUMO is just 0.5 eV away from
E$_{F}$, which is 0.3 eV closer than that of the perturbed HOMO.
Note that this finding is different from the microscopic pictures of
other existing molecular junctions based on photochromic DTE and DTC
switching molecules,\cite{PRB05min,CPL05masakazu}
azobenzene,\cite{PRL04jun} and quintuple bond {[}PhCrCrPh]
molecules,\cite{JCP06jing} whose transport properties are prevailed
by the transmission peak contributed by the perturbed HOMO.

Yet contradictorily, the spatial profiles shown in Fig.3 (b) of the
perturbed LUMO strongly localizes at the central switching unit with
open configuration. This leads to no appreciable transmission peaks
in the wide bias window (i.e. from -1.5 to 1.7 eV). The significant
transmission peak at -1.7 eV originates from the perturbed HOMO and
HOMO-1 (both are $\pi$ orbitals) for the open structure, however, it
is located too far away from the $E_{F}$. Here, comparing to the
closed case, we note that the position of the perturbed HOMO for the
open configuration is buried deeply below the E$_F$, which is
consistent with the previous theoretical results of DTC
molecules.\cite{PRB05min,CPL05masakazu} These theoretical findings
ensure us to conclude that the sharp contrast of the alignment of
the perturbed orbital energies with respect to the electrode Fermi
level and the shape of these perturbed frontier molecular orbitals
are the essential causes for the striking contrast in transport
properties through diarylethene molecular junctions with closed and
open configurations.

It should be pointed out that the number of transmission paths can
not account for the dramatic difference in conductivities of the
closed and open configurations since the eigenchannel analysis
indicates that there is a single eigenchannel for both cases within
a wide window (i.e. [-1.5, 1.5 eV]). According to the features
mentioned above of the calculated zero-bias transmission spectra
(Fig. 3), One can speculate that this type of molecular switch can
operate robustly in a pretty wide range of bias voltages with fairly
large on -off ratio. Additional currents through the diarylethene
molecular junctions with two different configurations at -2.0, -1.5,
1.5 and 2.0 V are also calculated. The on-off ratios of current are
predicted to be about two orders of magnitude. This suggests that
the bias voltage window of this kind of molecular switch (in
Taniguchi \emph{et al.}'s experiments\cite{JACS06masateru}) with
reasonably large on-off ratio is surprisingly wider than that of
other photo-sensitive molecules.\cite{PRB05min,PRL04zhang}

Experimentalist found that diarylethene molecular switch is
reversible when the molecules are sandwiched through aromatic
linkers.\cite{Adv06nathalie,JACS06masateru} Theoretical calculations
argued that whether it can be switched reversibly or not depending
on the molecule-electrode hybridization.\cite{PRB05min} The weak
interaction between molecule and electrode is required to facilitate
the desired reversible transition. According to these findings, the
reversible transition between the open and closed configurations in
this diarylethtene derivatives based molecular junction is highly
possible, since the molecules are sandwiched with phenyl linkers and
the molecule-electrode hybridization is weak.

\subsection{Substituting effect on diarylethene molecule}

Previous theoretical calculations focus on the end linking
groups,\cite{PRL04jun} no attempts so far have been made to examine
the side substituting effect on transport properties through the
diarylethene derivations. It is important to investigate the
conductance of the molecular junction, where six F atoms in the
peripheral of cyclopentene are substituted by H atoms. The
calculated transmission spectra for the H-substituted molecular
junction with the closed and open forms at zero bias voltage are
shown by black solid lines in Figures. 4(a) and 4(b), and two
corresponding I-V curves are presented in Fig. 2 with filled and
empty circles (linked by red dotted lines), respectively. The
current through the H-substituted molecular junction with closed
configuration is about half of that of the cyclopentene with six F
atoms in the peripheral. The reasons are summarized in the following
three points. (1) The replacement of H with F on the switching unit
results in the variation of band gaps. The energy gap of the
H-substituted diarylethene molecule is about 1.7 eV, while the gap
of diarylethene (F) molecule is about 1.3 eV. (2) The alignment of
the Fermi level is different for two systems. For the junction with
the H-substituted molecule, the peak coming from the perturbed HOMO
locates at -0.7 eV, which is closer to the Fermi level than the
perturbed LUMO transmission peak (at 1.0 eV). This result is
consistent with these previous theoretical studies on other DTC and
DTE derivations.\cite{PRB05min,CPL05masakazu} However, the Fermi
level lies close to the transmission contributed by the perturbed
LUMO for the diarylethene (F) molecular junction, as shown in Fig.
3. (3) The transport properties under small bias voltage are mainly
determined by the tail of the transmission peak coming from the
perturbed LUMO for the closed diarylethene (F) molecular junction.
However, the conductivity of the closed H-substituted one is
controlled the tail of transmission peak contributed by the
perturbed HOMO. Nonetheless, the light-controlled switching feature
is undoubtedly retained.

Experimental and theoretical results revealed that the visible
adsorption spectra changed when two S atoms of the switching unit
were substituted by O atoms.\cite{JOC88} Thus, the transmission
spectra of the molecular junction shown in Fig.1 (b) where two S
atoms in central switching unit are replaced by O atoms are also
calculated here, as shown in Fig. 4 with red dotted lines. Clearly,
the switching behavior does not depend sensitively on the
O-substituent. However, it should be pointed out that the positions
of significant transmission peaks obviously shift when compared to
Fig. 3. Particularly, the transmission peaks coming from the
perturbed HOMO and LUMO locates at -0.7 and 0.8 eV, respectively.
Again, the tail of perturbed HOMO transmission peak contributes
largely to the low bias electronic conductance.

\subsection{Effect of varing end anchoring atoms }

In general, the transport properties of molecular junctions depend
nontrivially on the end linking atoms.\cite{JACS04san,PRB04yong} Now
we turn to explore the effect of alternating end anchoring atoms.
The calculated transmission spectra at zero bias voltage are shown
in Figure 5. The black solid and red dotted lines stand for end Se-
and Te-anchored cases, respectively. It is clear that the main
characteristics of the transmission spectra are maintained and the
closed structure is undoubtedly more conductive. For the end
Se-anchored case, the energies of perturbed MO are quite close to
the data presented in Fig. 3 of the end S-anchored one.
Interestingly, clearly observable changes have been shown for the
end Te-anchored case. The transmission peaks originating from the
perturbed HOMO and LUMO for the molecular junction connecting to
gold electrodes through Te atoms locates at -1.0 (-0.8 for
S-anchored one) and 0.3 eV (0.5 for S-anchored one), respectively.
The very interesting finding of this study is that the switching
behavior of diarylethene derivatives based molecular junctions is
robust to vary end anchoring atoms from S to Se and Te.

To examine the sensitivity of results shown in Fig. 3 to small change
of the electrode-electrode distance, we compute the zero-bias transmission
spectra of diarylethene switches with the closed and open structures
as elongating and shortening electrode-electrode separation up to
0.3 {\AA}. We find that the transmission spectra experience little
change, and transport properties of this kind of diarylethene molecular
junction is not detectably sensitive to the electrode-electrode distance.
It indicates that this kind of light-controlled switching based on
diarylethene derivatives is stable as a molecular switching device.
Note that the transport behavior is described by the electron elastic
scattering theory in our calculations. The effect arising from the
electronic vibration and the accompanying heat dissipation on the
calculated on-off ratio can be neglected because of the remarkable
difference of the I-V curves.\cite{PRB05viljas}

\section{CONCLUSION}

In summary, we investigate the transport properties of the
diarylethene with closed and open structures using the NEGF combined
the DFT method. The zero-bias transmission function of two different
forms is strikingly distinctive. The open form lacks any significant
transmission peak within a wide energy window, while the closed
structure has two significant transmission peaks on the both sides
of the Fermi level. The electronic transport properties of the
molecular junction with closed structure under a small bias voltage
are mainly determined by the tail of the transmission peak
contributed unusually by the perturbed lowest perturbed unoccupied
molecular orbital. The calculated on-off ratio of currents between
the closed and open configurations is about two orders of magnitude,
which reproduces the essential features of the experimental measured
results. Moreover, although the alignments of the perturbed
molecular orbitals's energies with respect to the electrode's Fermi
level are not exactly the same, we find that the switching behavior
within a wide bias voltage window is extremely robust to both
substituting F or S for H or O and varying end anchoring atoms from
S to Se and Te.

\section*{ACKNOWLEDGMENTS}

This work was partially supported by the National Natural Science
Foundation of China under Grants 10674121, 10574119, 50121202, and
20533030, by National Key Basic Research Program under Grant No.
2006CB922004, by the USTC-HP HPC project, and by the SCCAS and
Shanghai Supercomputer Center.Work at NTU is supported in part by
A{*}STAR SERC grant (No. 0521170032).

\newpage
\begin{figure}[htbp]

\caption{(Color online) (a) The diarylethene derivative in closed
and open configurations. (b) A schematic of the switching junction.
Diarylethene molecules are sandwiched between two Au (111) surfaces,
and two S anchoring atoms are located at the hollow site. The
vertical blue line denotes the interface between the scattering
region and the left or right gold electrode.}

\caption{(Color online) The calculated current-voltage
characteristics of the diarylethene and its derivative molecular
junctions with two different configurations. The triangles linking
with black solid lines are for the diarylethene molecular junction,
while the circles linking with short red dotted lines stand for the
junction where six F atoms in the peripheral of cyclopentene are
substituted by H atoms. The filled (empty) symbols correspond to the
closed (open) structures. The inset below right is the I-V curve for
the open structures (with F and H atoms in the peripheral,
respectively) in small scale for clarity.}

\caption{(Color online) The zero-bias voltage transmission spectra
versus the energy E-E$_{F}$ of diarylethene molecular junctions with
the closed (a) and open (b) configurations. Here, E$_{F}$ is the
Fermi level of electrodes. The red short vertical lines stand for
the positions of MPSH molecular energy levels. The spatial
distributions of the perturbed HOMOs and LUMOs are inserted in the
figure, and placed at the right and left sides of the E$_{F}$,
respectively. }

\caption{(Color online) The calculated transmission spectra versus
the energy E-E$_{F}$ at zero-bias voltage for diarylethene molecular
junctions with closed (a) and open (b) forms, respectively. One case
is that six F atoms in the peripheral of central cyclopentene are
substituted by H atoms (with black solid lines); the other is that
two S atoms are replaced by O atoms (with red dotted lines). The red
short vertical lines stand for the positions of MPSH molecular
energy levels.}

\caption{(Color online) The calculated transmission spectra for
diarylethene molecular junctions with closed (a) and open (b)
structures. The black solid and red dotted lines stand for the end
anchoring Se and Te atoms, respectively. Here, the red short
vertical lines stand for the positions of MPSH molecular energy
levels. }
\end{figure}

\clearpage
\begin{figure}[htbp]
\includegraphics[width=10cm]{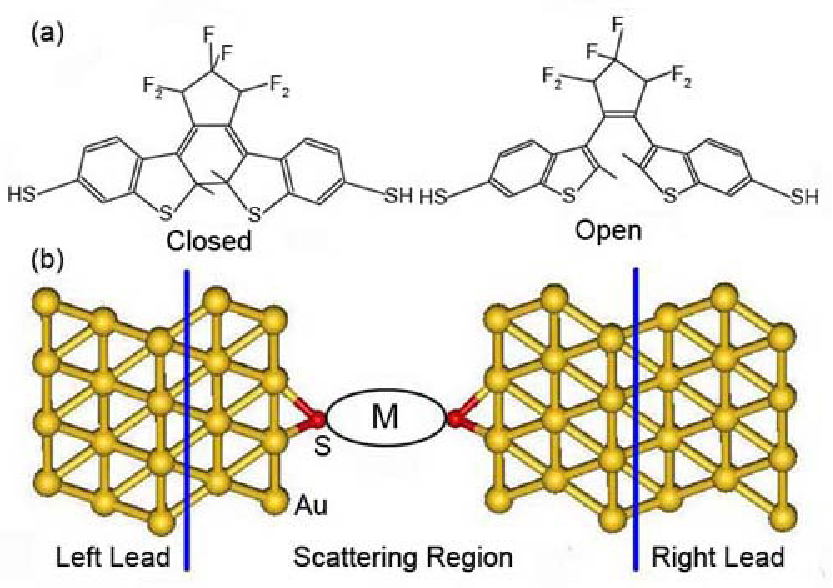}
\begin{center} \vspace{5cm} Fig.1 of Huang \emph{et al.} \end{center}
\end{figure}

\clearpage
\begin{figure}[htbp]
\includegraphics[width=10cm]{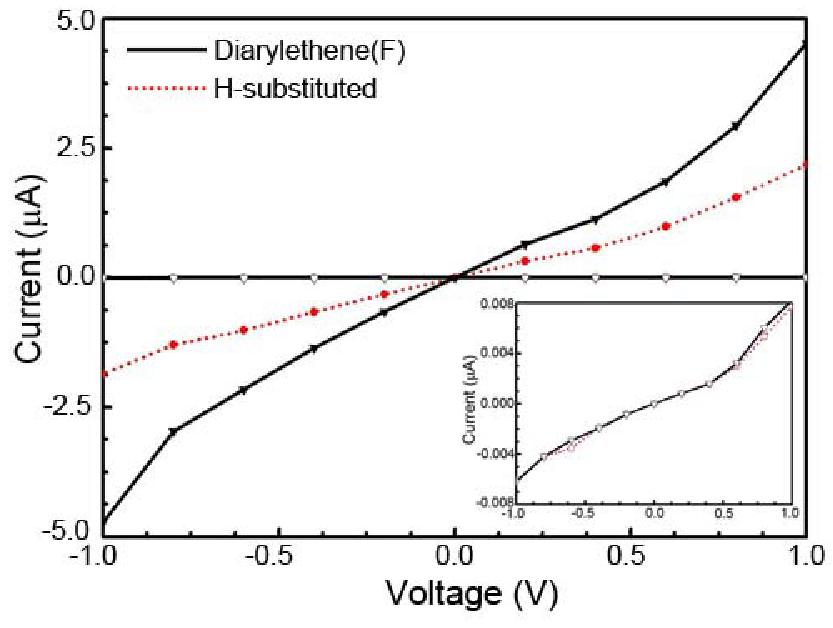}
\begin{center} \vspace{5cm} Fig.2 of Huang \emph{et al.} \end{center}
\end{figure}

\clearpage
\begin{figure}[htbp]
\includegraphics[width=10cm]{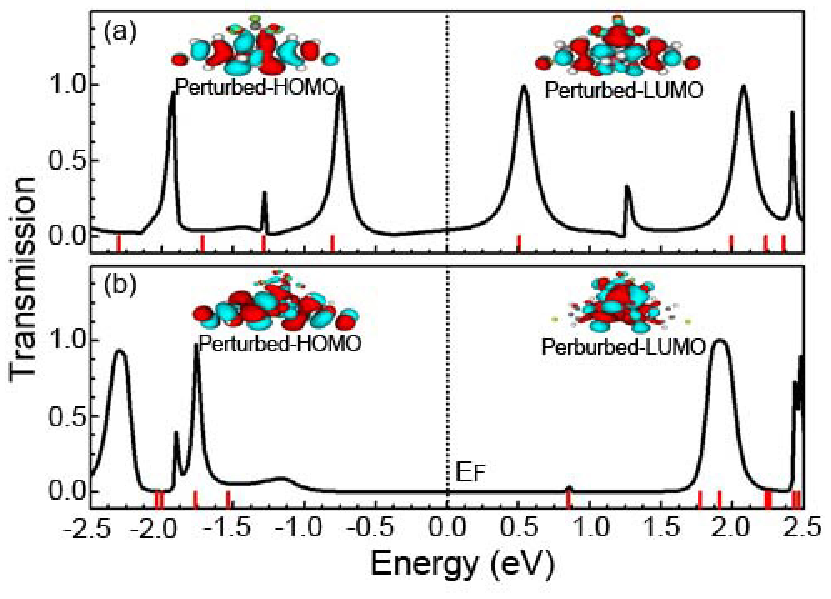}
\begin{center} \vspace{5cm} Fig.3 of Huang \emph{et al.} \end{center}
\end{figure}

\clearpage
\begin{figure}[htbp]
\includegraphics[width=10cm]{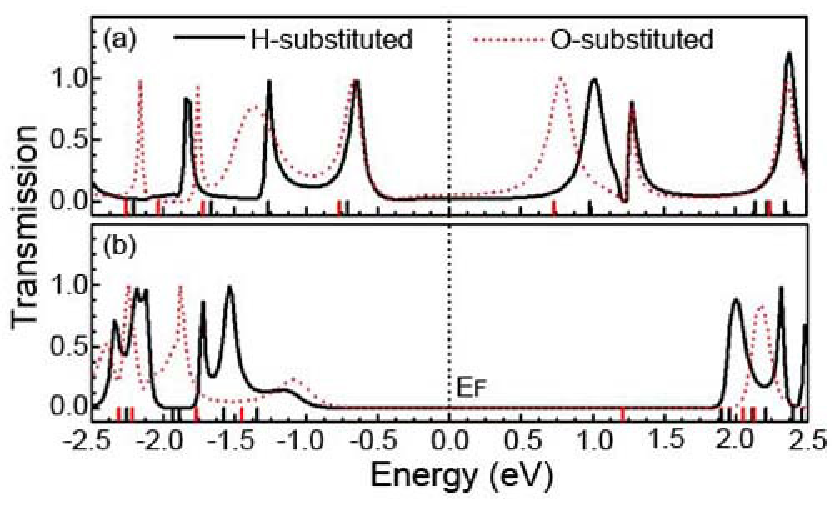}
\begin{center} \vspace{5cm} Fig.4 of Huang \emph{et al.} \end{center}
\end{figure}

\clearpage
\begin{figure}[htbp]
\includegraphics[width=10cm]{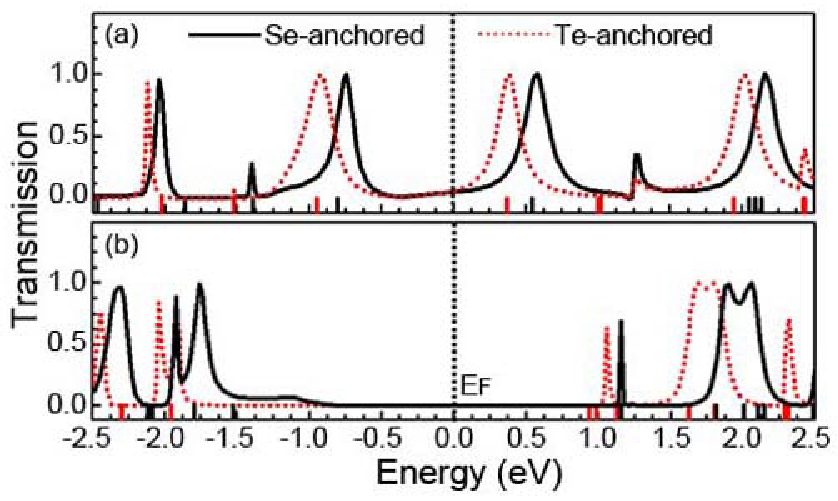}
\begin{center} \vspace{5cm} Fig.5 of Huang \emph{et al.} \end{center}
\end{figure}


\begin{references}
\bibitem{adsorbed} We have performed DFT calculations for the
diarylethene molecule adsorbing on the hollow, bridge, and atop
sites. The calculated results show that the molecule prefers to the
hollow site. Its adsorption energy is lower about 0.2 and 0.6 eV
than that of the bridge and atop adsorption configuration,
respectively.
\bibitem{Supply} Single-zeta plus polarization (SZP) basis set for
Au atoms and double zeta plus polarization (DZP) basis set for other
atoms are adopted. Test calculations show that the very similar results
are obtained by using DZP basis set for all atoms.
\end{references}
\end{document}